\documentclass[aps,prl,twocolumn,amsmath,amssymb,superscriptaddress,longbibliography]{revtex4-1}
\usepackage[english]{babel}
\usepackage{xcolor}
\usepackage{amssymb} 
\usepackage{dcolumn}
\usepackage{bm}
\usepackage{graphicx}
\usepackage{amsmath}
\usepackage{braket}

\usepackage{graphicx}        % standard LaTeX graphics tool

\graphicspath{{pict/}{}}

\usepackage[normalem]{ulem}

\usepackage{dcolumn}%%%%%new
\usepackage{bm}
\usepackage[pdfstartview=FitH, CJKbookmarks=true, bookmarksnumbered=true, bookmarksopen=true, colorlinks=true, pdfborder=001, citecolor=blue, linkcolor=blue, urlcolor=blue, linktocpage=true] {hyperref}

\makeatletter

\begin{document}

\date{\today}

%\title{Deterministic Single to Bundle Photon Emissions in Cavity-Coupled Reconfigurable Atom Arrays}

%\title{Single to Bundle Photon Emissions in Cavity-Coupled Reconfigurable Atom Arrays}

\title{Tunable Single- and Multiphoton Bundles in Cavity-Coupled Atomic Arrays}

\author{Geng Zhao}
\thanks{These authors contributed equally to this work.}
\affiliation{Guangdong Provincial Key Laboratory of Quantum Metrology and Sensing $\&$ School of Physics and Astronomy, Sun Yat-Sen University, Zhuhai 519082, China}

\author{Yun Chen}
\thanks{These authors contributed equally to this work.}
\affiliation{Guangdong Provincial Key Laboratory of Quantum Metrology and Sensing $\&$ School of Physics and Astronomy, Sun Yat-Sen University, Zhuhai 519082, China}

\author{Jiayuang Zhang}
\affiliation{Guangdong Provincial Key Laboratory of Quantum Metrology and Sensing $\&$ School of Physics and Astronomy, Sun Yat-Sen University, Zhuhai 519082, China}

\author{Jing Tang}
\email{jingtang@gdut.edu.cn}
\affiliation{School of Physics and Optoelectronic Engineering, Guangdong University of Technology, Guangzhou 510006, China}
\affiliation{Guangdong Provincial Key Laboratory of Sensing Physics and System Integration Applications, Guangdong University of Technology, Guangzhou, 510006, China}

\author{Yuangang Deng}
\email{dengyg3@mail.sysu.edu.cn}
\affiliation{Guangdong Provincial Key Laboratory of Quantum Metrology and Sensing $\&$ School of Physics and Astronomy, Sun Yat-Sen University, Zhuhai 519082, China}

%\affiliation{School of Physics and Astronomy, Sun Yat-Sen University, Zhuhai 519082, China}
%\affiliation{Guangdong Provincial Key Laboratory of Quantum Metrology and Sensing, Sun Yat-Sen University, Zhuhai 519082, China}

\begin{abstract}
We propose an experimentally accessible scheme for realizing tunable nonclassical light in cavity-coupled reconfigurable atomic arrays. By coherently controlling the collective interference phase, the system switches from single-photon blockade to high-purity multiphoton bundle emission, unveiling a hierarchical structure of photon correlations dictated by atom-number parity and cavity detuning. The scaling of photon population identifies the transition between superradiant and subradiant regimes, while parity- and phase-dependent spin correlations elucidate the microscopic interference processes enabling coherent multiphoton generation. This work establishes a unified framework connecting cooperative atomic interactions to controllable nonclassical photon statistics and introduces a distinct interference-enabled mechanism that provides a practical route toward high-fidelity multiphoton sources in scalable cavity QEDs. 
\end{abstract}

\maketitle
{\em Introduction}.---The realization of tunable light-matter interfaces lies at the heart of quantum science, providing essential building blocks for quantum information and precision metrology~\cite{RevModPhys.90.031002,RevModPhys.90.045005,li2023improving,PhysRevLett.132.060801,lkrt-lvng}. Cavity quantum electrodynamics (QEDs) offers a natural platform for exploring strongly coupled systems, enabling engineering highly correlated nonclassical states with tailored coherence properties. Significant progress has been achieved in developing strongly nonlinearities for realizing photon blockade~ \cite{birnbaum2005photon,aoki2006observation,srinivasan2007linear,dayan2008photon,liew2010single,PhysRevLett.118.133604,PhysRevLett.134.183601,PhysRevLett.134.013602} and multiphoton bundle states~\cite{C2014Emitters,PhysRevLett.127.073602,liu2023deterministic,PhysRevLett.133.043601}, establishing a pathway toward scalable quantum networks~\cite{ritter2012elementary,kimble2008quantum,redjem2023all} and nondestructive quantum measurements~\cite{niemietz2021nondestructive,duan2004scalable,reiserer2013nondestructive}. Concurrently, optical tweezer arrays of neutral atoms and polar molecules have emerged as a versatile platform combining single-particle addressability, long coherence times, and programmable long-range interactions~\cite{ebadi2021quantum,2012Quantum,doi:10.1126/science.aah3778,anderegg2019optical,cairncross2021assembly}. These arrays have enabled advances in quantum simulation of strongly correlated many-body physics~\cite{bernien2017probing,zhang2017observation,de2019observation,yue2025observing}, scalable quantum computing~\cite{demille2002quantum,browaeys2020many,kaufman2021quantum,PhysRevLett.129.123201}, deterministic entanglement~\cite{wilk2010entanglement,holland2023demand,bao2023dipolar,schine2022long}, and precision quantum sensing~\cite{marciniak2022optimal,PhysRevLett.123.260505,PRXQuantum.5.010311}. 

\begin{figure}[ptb]
\includegraphics[width=0.9\columnwidth]{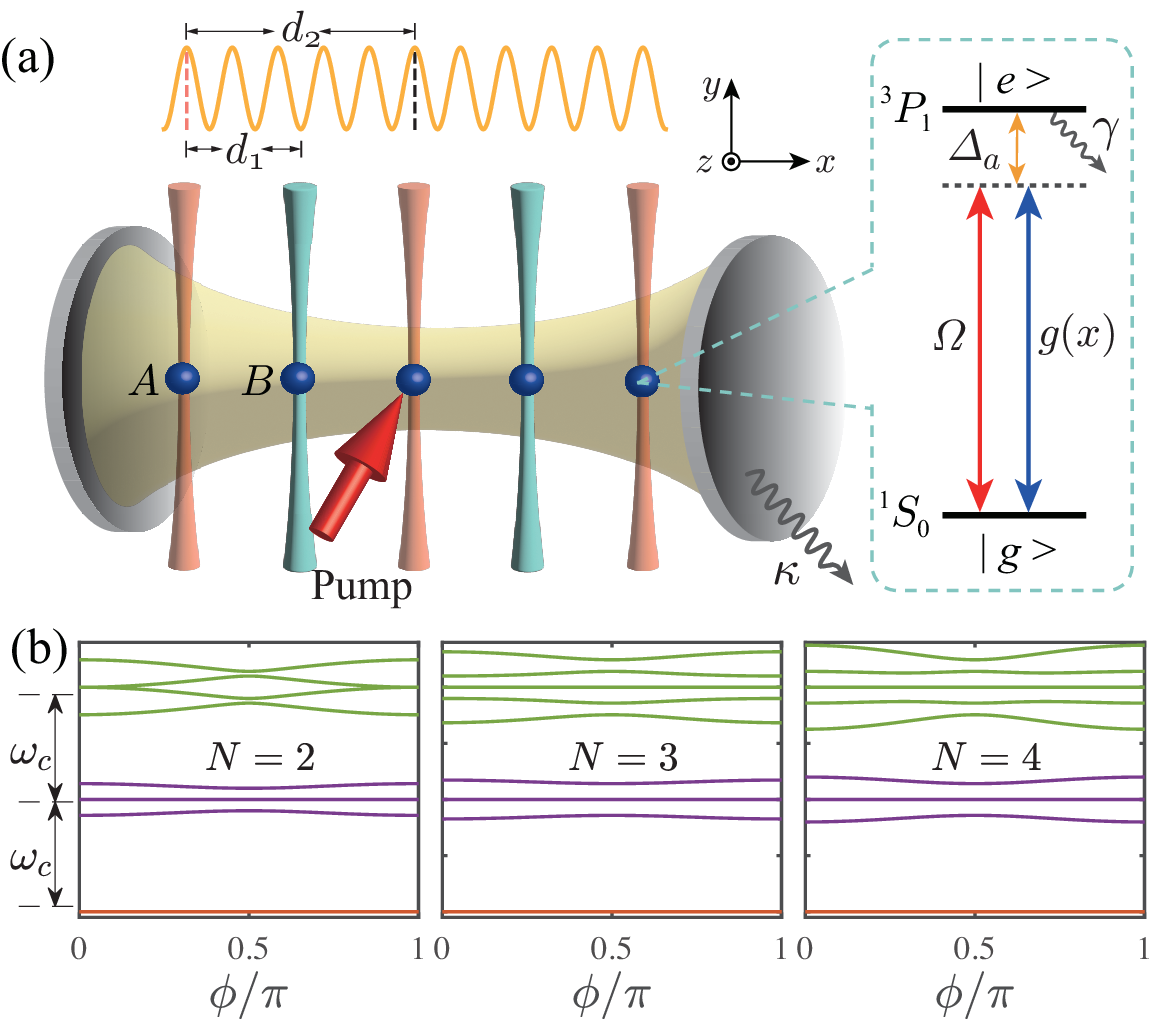} \caption{\label{fig_model}
(a) Schematic of cavity-coupled reconfigurable atomic array, highlighting dipole-forbidden $^1S_0 \leftrightarrow {^3P_1}$ transition of $^{88}$Sr.  (b) Phase $\phi$ dependence of anharmonic energy spectrum for different atom numbers $N$.}
\end{figure}

Integrating such atomic arrays into high-finesse optical cavities introduces cavity-mediated infinite-range interactions~\cite{PhysRevLett.114.023601,PhysRevLett.116.223001,PhysRevLett.118.210503,PhysRevLett.114.023602,dhordjevic2021entanglement,PhysRevLett.128.083201}, opening new avenues to explore nonequilibrium quantum phases of light and matter. Recent experiments with cavity-coupled atom arrays have revealed cooperative radiation phenomena, including superradiance and subradiance~\cite{PhysRevLett.131.253603,PhysRevLett.130.173601,masson2020many,PhysRevLett.133.106901,PhysRevLett.133.243401},  arising from phase-dependent interference determined by the spatial arrangement of emitters. Controllable switching between these regimes provides an unprecedented handle to engineer collective emission, decoherence, and photon correlations~\cite{rui2020subradiant}. Despite these advances, the emergence of nonclassical multiphoton bundle states across radiance transitions remains largely unexplored. Establishing a microscopic connection between collective atomic correlations and photon-number hierarchy is crucial for bridging cooperative atom-light physics and nonclassical photon generation.

In this Letter, we present an experimental scheme to achieve controllable nonclassical light emission in cavity-coupled atomic arrays. By tuning the collective interference phase between neighboring atoms, the system undergoes a transition from superradiant to subradiant regimes, accompanied by the emergence of high-quality multiphoton bundle state enabled by interference-suppressed single-photon excitation. We show that the scaling of photon population with atom number directly reflects the collective radiance transition, while parity- and phase-dependent spin correlations reveal the microscopic origin of bunching multiphoton generation. In contrast to recent studies focusing on collective emission in the vacuum-Rabi splitting or far-dispersive regimes~\cite{PhysRevLett.131.253603,PhysRevLett.130.173601,masson2020many,PhysRevLett.133.106901,PhysRevLett.133.243401}, our study identifies a radiance-controlled quantum switch between single-photon and multiphoton bundle sources operating in the atom-cavity resonant regime. These results establish a unified framework linking collective light-matter correlations to nonclassical photon statistics. Remarkably, unlike previously explored routes to bundle state---such as Mollow physics~\cite{C2014Emitters}, deterministic parametric down-conversion~\cite{PhysRevLett.117.203602}, n-phonon resonance processes~\cite{deng2021motional}, and parity-symmetry-protected multiphoton process~\cite{PhysRevLett.127.073602}, our cavity-coupled reconfigurable atom arrays provide a practical route  toward on-demand multiphoton emitters for quantum technologies and precision metrology~\cite{neuzner2016interference}. 

{\em Model}.---We consider atom arrays assembled with an one-dimensional optical tweezer array placed inside a high-finesse Fabry-P\'erot cavity [Fig.~\ref{fig_model}(a)]~\cite{PhysRevLett.131.253603, PhysRevLett.130.173601}. As a representative system, we focus onalkaline-earth  atoms such as $^{88}$Sr, whose positions can be precisely controlled~\cite{PhysRevLett.118.263601}. The relevant optical transition couples the ground state of singlet $^1S_0=|g\rangle$ to long-lived excited state of triplet $^3P_1=|e\rangle$, interacting with cavity mode along $x$ with single-atom-cavity coupling $g$ and a transverse pump along $y$ with Rabi frequency $\Omega$. Owing to its dipole-forbidden nature, this transition features a narrow linewidth, with a spontaneous decay rate $\gamma=(2\pi)7.5$ kHz at $\lambda=689$ nm~\cite{PhysRevLett.118.263601}. Distinct from uniform arrays, the setup implements a reconfigurable two-sublattice geometry with tunable lattice constants $d_1$ and $d_2$. The relative phase $\phi=kd_1$ (with wave vector $k=2\pi/\lambda$) controls constructive or destructive interference, while odd-site atoms are prepared at integer-wavelength separations ($d_2=5\lambda$). By tuning $d_1$, interference pathways can be programmed, enabling the generation of nonclassical photon states. For $N\leq60$ atoms, the array length ($\leq103~\mu$m) remains well within the cavity Rayleigh range ($\sim1~\text{mm}$), ensuring uniform coupling. This reconfigurable architecture offers a versatile platform for engineering controllable quantum sources spanning single-photon to multiphoton emission.

For an array of $N$ spatially fixed two-level atoms, Rayleigh scattering from pump field into single-mode cavity can be mapped onto an effective two-component Tavis-Cummings model (TCM)
\begin{align} \label{Ham}
{\cal {\hat H}}=&\Delta_c \hat{a}^\dagger \hat{a}+\Delta_a (\hat{J}^z_{A}+\hat{J}^z_{B})+\Omega (\hat{J}_{A}^x+\hat{J}_{B}^x) \nonumber\\
&+ g\hat{a}^\dagger (\hat{J}^-_{A} + \cos \phi \hat{J}^-_{B}) + g\hat{a} (\hat{J}^+_{A} + \cos \phi \hat{J}^+_{B}),
\end{align}
where $\hat{a}$ is the cavity annihilation operator, $\Delta_c $ and $\Delta_a$ are the cavity- and atom-pump detunings, respectively. The collective spin operators $\hat{J}_{A,B}^{x,y,z}=\frac{1}{2}\sum^{N_{A,B}}_{j}\hat{\sigma}^{x,y,z}_j$ describe atoms in sublattices $A$ and $B$, with $\hat{J}^{\pm}_{A,B}=\hat{J}_{A,B}^{x} \pm i\hat{J}_{A,B}^{y}$, $N_{A(B)}$ being atom number in each sublattice, and $\hat{\sigma}^{x,y,z}$ being Pauli matrices. The light-matter interaction encodes phase-dependent coupling between cavity and sublattices, enabling interference-controlled cavity emission. Notably, non-interacting neutral atoms at different sites are coupled by the infinite-range cavity via Stokes processes with a photon annihilation accompanied by atomic excitation. Neglecting the weak transverse drive ($\Omega=0$), the Hamiltonian (\ref{Ham}) exhibits a continuous $U(1)$ symmetry generated by $\mathcal{R}_{\theta}=\exp(i\theta \hat{N}_e)$, with $\hat{N}_e = \hat{a}^\dagger \hat{a} + \hat{J}_A^z + \hat{J}_B^z$ the total excitation number. This symmetry commutes with the Hamiltonian, $[\mathcal{R}_{\theta},{\cal {\hat H}}]=0$, and its spontaneous breaking signals the optical switching from normal to superradiant phase~\cite{tang2024unveiling}.

Taking into account all dissipations, nonclassical photon emission is captured by solving the master equation 
 \begin{equation}\label{master equation}%
{ \frac{d\rho}{dt}}= -i [\hat{\cal H}, {\rho}] + {\kappa} \mathcal
{\cal{D}}[\hat{a}]\rho + \frac{\gamma}{N_A} \mathcal
{\cal{D}}[\hat{J}_{A}^-]\rho +  \frac{\gamma}{N_B} \mathcal
{\cal{D}}[\hat{J}_{B}^-]\rho,
\end{equation}
where $\rho$ is the density matrix of cavity-coupled atomic arrays, $\mathcal {D}[\hat{o}]\rho=\hat{o} {\rho} \hat{o}^\dag - (\hat{o}^\dag \hat{o}{\rho} + {\rho}\hat{o}^\dag \hat{o})/2$ denotes standard Lindblad-type dissipation, and $\kappa=(2\pi)75$ kHz is the cavity decay rate~\cite{941q-5sdq,kongkhambut2022observation}. In contrast to alkaline-metal atom arrays with $\gamma/\kappa \gg 1$, the small ratio $\gamma/\kappa =0.1$ is crucial for realizing high-fidelity multiphoton-bundle emission. In numerical simulations, we focus on atom-photon resonance, $\Delta_c=\Delta_a=\Delta$, single atom-cavity coupling $g/\kappa=10$, and Rabi frequency $\Omega/\kappa=0.2$.
 
To explore the quantum statistics, we introduce the generalized $k$th-order correlation function~\cite{PhysRevResearch.6.033247} 
\begin{align} 
g_n^{(k)}(\tau_1,\ldots,\tau_k)=\frac{\left\langle \prod_{i=1}^k\left[\hat{a}^{\dagger }(\tau_i)\right]^n \prod_{i=1}^k\left[\hat{a}(\tau_i)\right]^n\right\rangle}{\prod_{i=1}^k\left\langle \left[\hat{a}^{\dagger }(\tau_i)\right]^n\left[\hat{a}(\tau_i)\right]^n \right\rangle}, \nonumber
\label{g220}%
\end{align} 
with $\tau_1\leq...\leq\tau_k$. This function captures nonclassical emission ranging from isolated photons to $n$-photon bundles. Explicitly, $g_1^{(k)}$ reduces to the conventional $k$th-order correlation function for isolated photons. Single photon blockade is identified by $g^{(2)}_1(0)<1$ and $g_1^{(2)}(0)<g_1^{(2)}(\tau)$, which indicates sub-Poissonian statistics and photon antibunching, respectively. By contrast, the criteria for $n$-photon bundle emission is $g_1^{(2)}(0)>g_1^{(2)}(\tau)$ and  $g_n^{(2)}(0)<g_n^{(2)}(\tau)$, ensuring photon bunching within each bundle and antibunching between separated bundles~\cite{PhysRevLett.117.203602,C2014Emitters, deng2021motional}.

{\em Energy spectrum}.---To reveal the mechanism of nonclassical photon emission, we first analyze the energy spectrum of cavity-coupled reconfigurable atomic array by diagonalizing the Hamiltonian (\ref{Ham}). In the single-excitation subspace, the eigenenergies of three branches satisfy: $E_{1,\pm}=\Delta\pm g\sqrt{(N_A+N_B\cos^2 \phi)}$ and $E_{1,0}=\Delta$, with corresponding eigenstates
\begin{eqnarray}
	\label{eq_eigenvalue_single_ex}
	&&|\Psi_{1,\pm} \rangle=\frac{1}{\sqrt{2}} [|1,0 \rangle\pm (\beta_A|0,1_A\rangle+
\beta_B|0,1_B \rangle)],\nonumber\\
	&&|\Psi_{1,0} \rangle=-\beta_B|0,1_A\rangle+\beta_A|0,1_B \rangle),
	\end{eqnarray}
where $|n,0\rangle = |n,-\frac{N_A}{2},-\frac{N_B}{2}\rangle$ with $n$ being the photon number, $|n-1,1_{A} \rangle = |n-1,1-\frac{N_{A}}{2},-\frac{N_B}{2}\rangle$, and $|n-1,1_{B} \rangle = |n-1,-\frac{N_{A}}{2},1-\frac{N_B}{2}\rangle$ are introduced by shorthand notations. Here $\beta_A={\sqrt{N_A}}/{\sqrt{N_A+N_B\cos^2\phi}}$ and $\beta_B={\sqrt{N_B}\cos\phi}/{\sqrt{N_A+N_B\cos^2\phi}}$. 

\begin{figure}[ptb]
\includegraphics[width=0.46\textwidth]{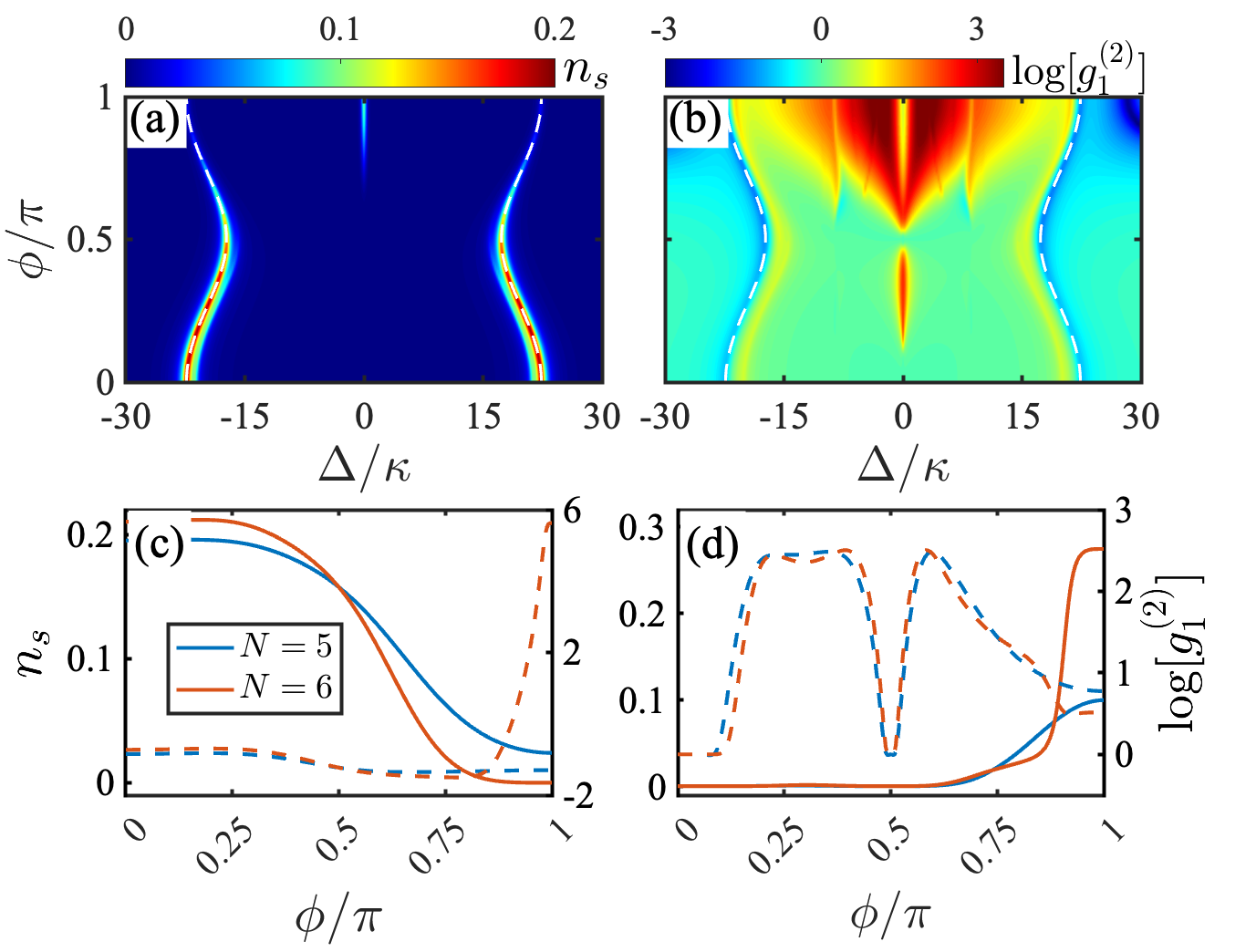} 
\caption{\label{fig2_N=45} Distributions of (a) $n_s$ and (b) $g^{(2)}_1(0)$ on the $\Delta-\phi$ plane for $N=5$. The white dashed curves indicate the analytic dressed-state splitting at $\Delta=\Delta_{1,\pm}$. The phase $\phi$ dependence of  $n_s$ (solid line) and $g^{(2)}_1(0)$ (dashed line) for (c) $\Delta=|\Delta_{1,\pm}|$ and (d) $\Delta=0$ at $N=5$ and 6, respectively. Color shading encodes the magnitude of $n_s$ and $g^{(2)}_1(0)$.}
\end{figure}

For $\phi=0$, the spectrum contains two bright polaritons $|\Psi_{1,\pm} \rangle$ and one dark state $|\Psi_{1,0} \rangle$, corresponding to  symmetric and anti-symmetric superpositions of $A$- and $B$-sublattice excitations. At $\phi=\pi$, this symmetry is reversed, while for arbitrary $\phi$, the dark state persists with vanishing photon number at the single excitation subspace. Remarkably, even in weak-drive regime $\Omega/g \ll 1$, photon emission still occurs through the middle branch ($\Delta=0$), indicating that multiphoton processes beyond the single-excitation manifold must be involved.

To reveal this mechanism, we calculate the full multiphoton spectrum, as detailed in Supplementary Materials~\cite{SM}. For two atoms, the analytic $n$-photon dressed states associated with the middle branch at $\phi=\pi$ take the form ($n \geq 2$)
  \begin{eqnarray}\label{multiphoton}
|\Psi_{n,0_+}\rangle&&=\frac{1}{\sqrt{2}}(|n-1,1_A \rangle+|n-1,1_B \rangle),\nonumber\\
|\Psi_{n,0_-}\rangle&&=\frac{1}{\sqrt{2}}(|n,0 \rangle+|n-2,2\rangle),
\end{eqnarray}
with $|n-2,2 \rangle = |n-2,1-\frac{N_A}{2},1-\frac{N_B}{2}\rangle$. Crucially, both $|\Psi_{n,0_{\pm}}\rangle$ are symmetric across sublattices and remain degenerate at zero energy enabling multiphoton emission via destructive interference completely suppressing single-photon excitation. In particular, two-photon bundle emission arises from super-Rabi oscillations of $\left| 2, 0\right\rangle \leftrightarrow \left| 0, 2 \right\rangle$, and analogous multiphoton super-Rabi processes persist for $N \geq 3$. In contrast to earlier studies where destructive interference suppressed emission at $\phi=\pi$~\cite{PhysRevLett.130.173601}, here the symmetric $n$th-order dressed states dominate when atoms rather than the cavity are directly driven, giving rise to nonclassical multiphoton emission. Notably, the middle-branch bundle emission is entirely absent for $\phi=0$, consistent with quantum mechanical picture of collective scattering in cavity-coupled atom arrays~\cite{SM}.

Figure~\ref{fig_model}(b) illustrates $\phi$-dependent anharmonic energy spectrum for different total atom numbers $N$. In the single-excitation manifold, three helicity branches consistently appear, in agreement with the analytic expressions in Eq.~(\ref{eq_eigenvalue_single_ex}). Beyond this regime, the doubly excited spectrum becomes increasingly structured: it exhibits four, five, and six (including two degenerate) helicity branches for $N=2$, $N=3$, and $N\geq4$, respectively. Strikingly, both single- and multiphoton excitations remain resonant for middle helicity at $\Delta=0$, thereby opening a robust channel for generating multiphoton states that manifest as nonclassical photon bunching.

\begin{figure}[ptb]
\includegraphics[width=0.46\textwidth]{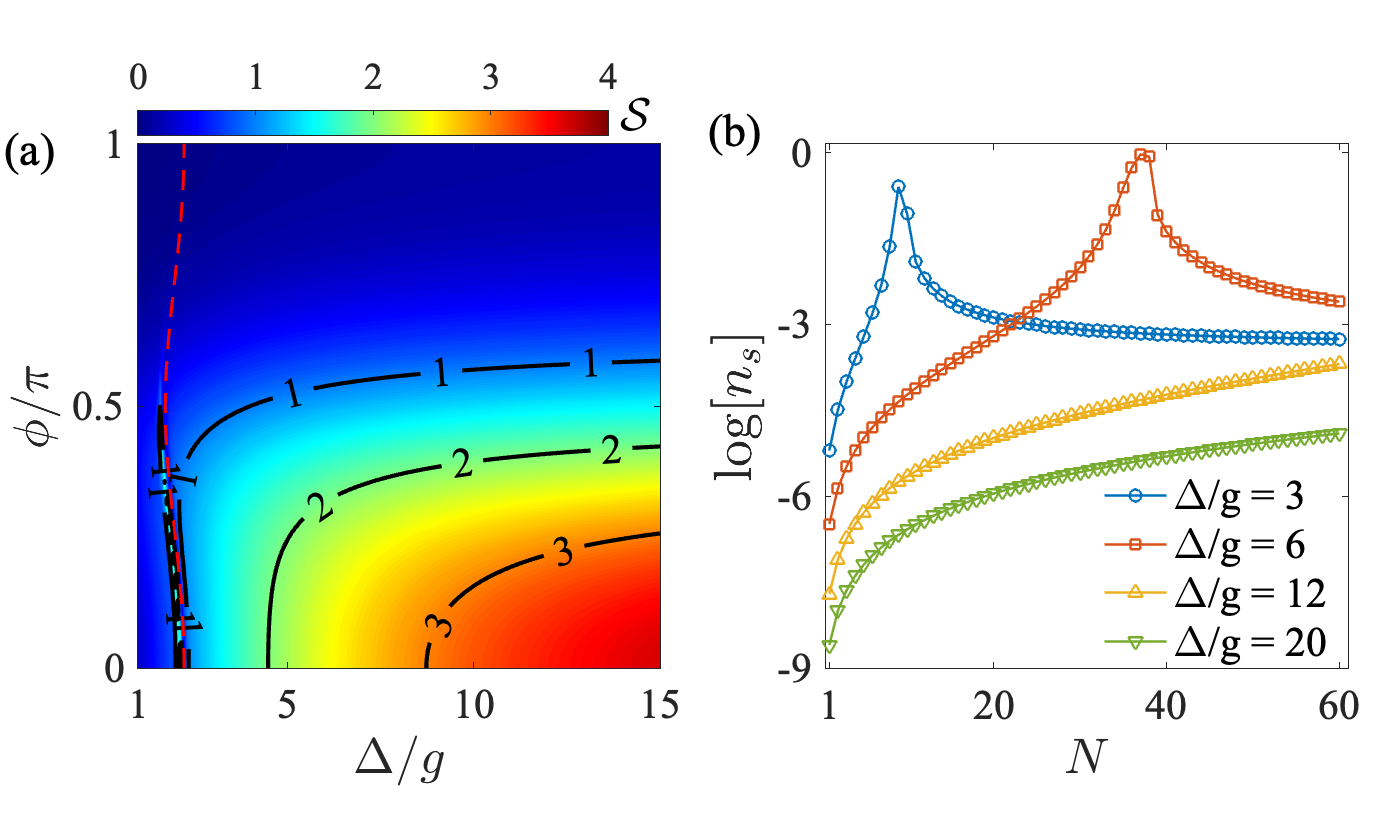} 
\caption{\label{fig4_scaling}(a) ${\cal S}$ as functions of $\Delta$ and $\phi$ for $N=5$. Solid lines are contour lines and the white dashed line shows vacuum Rabi splitting $\Delta_{1+}$ at blue sideband. (b) $n_s$ versus $N$ for $\phi=0$ and $\Delta/g=3,\,6,\,12,\, {\rm and}\, 20$. The black dashed line serves as a guide for $n_s\sim N^2$.} 
\end{figure}

{\em Subradiance to superradiance}.---Figures \ref{fig2_N=45}(a) and \ref{fig2_N=45}(b) display the steady-state photon number $n_s$ and second-order correlation function $g^{(2)}_1(0)$ as functions of detuning $\Delta$ and phase $\phi$ for $N=5$. The dominant photon emission appears at the single-photon resonances $\Delta_{1\pm}=\pm g\sqrt{N_A+N_B \cos^2\phi}$, which correspond to the bright polariton states $|\Psi_{1,\pm}\rangle$. These phase-dependent splittings $\Delta_{1\pm}$ represent an effective vacuum Rabi splitting, consistent with recent observations in cavity-coupled atomic arrays~\cite{PhysRevLett.130.173601}, and exhibit strong sensitivity to both $N$ and $\phi$. As shown in Fig.~\ref{fig2_N=45}(c), photon number displays a  pronounced contrast between constructive ($\phi=0$) and destructive ($\phi=\pi$) interference, signaling the superradiant to subradiant transition.

At red and blue sidebands, superradiant emission exhibits photon blockade characterized by $g^{2}_1(0)<1$. Notably, the blockade strength is not monotonic in $\phi$: the highest single-photon purity emerges for phase satisfying $\phi/\pi>0.5$. Furthermore, the optimal phase depends sensitively on both parity and magnitude of $N$~\cite{SM}, revealing the subtle influence of array symmetry on the radiative response. In contrast, at the middle branch ($\Delta=0$), the multiphoton dressed states with different excitation numbers $n$ become simultaneously resonant [see Eq.~(\ref{multiphoton})]. Consequently, the cavity output exhibits strong bunching with $g^{2}_1(0)>1$, while $n_s$ increases monotonically with $\phi$ [Fig.~\ref{fig2_N=45}(d)]. Remarkably, $n_s(\Delta=0, \phi=\pi)$ will reach a peak, regardless  of whether $N$ is even or odd, in sharp contrast to the bright sideband branches at $\Delta=|\Delta_{1\pm}|$. By symmetry, $n_s(\Delta=0, \phi=0)$ vanishes, since middle-branch state is antisymmetric at $\phi=0$ but becomes symmetric at $\phi/\pi=1$ with respective to two sublattices. 

To proceed further, we study the steady-state photon emission scales with total atom number $N$ in cavity-coupled reconfigurable arrays. Within the semiclassical treatment, the steady-state cavity population is
\begin{equation}
\label{steady_state}
n_s=\left|\frac{g\Omega(N_A  + N_B \cos\phi)}{\Delta(\Delta-i\kappa)-g^2(N_A +N_B \cos^2\phi)}\right|^2.
\end{equation}
At single-photon resonance $\Delta=\Delta_{1\pm}$, we obtain $n_s={N\Omega^2}/{\kappa^2}$ for $\phi=0$, revealing linear $N$ scaling of bright states $|\Psi_{1,\pm}\rangle$. By contrast, photon emission shows a strong parity effect at $\phi=\pi$: for even $N$ destructive interference completely suppresses cavity emission ($n_s=0$), whereas for odd $N$ we obtain $n_s=\Omega^2/(N\kappa^2)$, exhibiting an inverse $1/N$ scaling. For middle branch at $\Delta=0$ and $\phi=\pi$, the parity dependence persists: $n_s=\Omega^2/(gN)^2$ for even $N$, while it vanishes identically for odd $N$. In the dispersive regime $\Delta/g\gg1$, the photon number saturates to $n_s\sim (g\Omega/\Delta)^2$ for odd $N$ at $\phi=\pi$, while for $\phi=0$ it recovers the parity-independent superradiant scaling $n_s\sim (g\Omega N/\Delta)^2$. These results stem from mean-field analysis that neglects quantum fluctuations and correlations. The exact scaling behavior requires solving full master equation [Eq.~(\ref{master equation})].

\begin{figure}[ptb]
\includegraphics[width=0.46\textwidth]{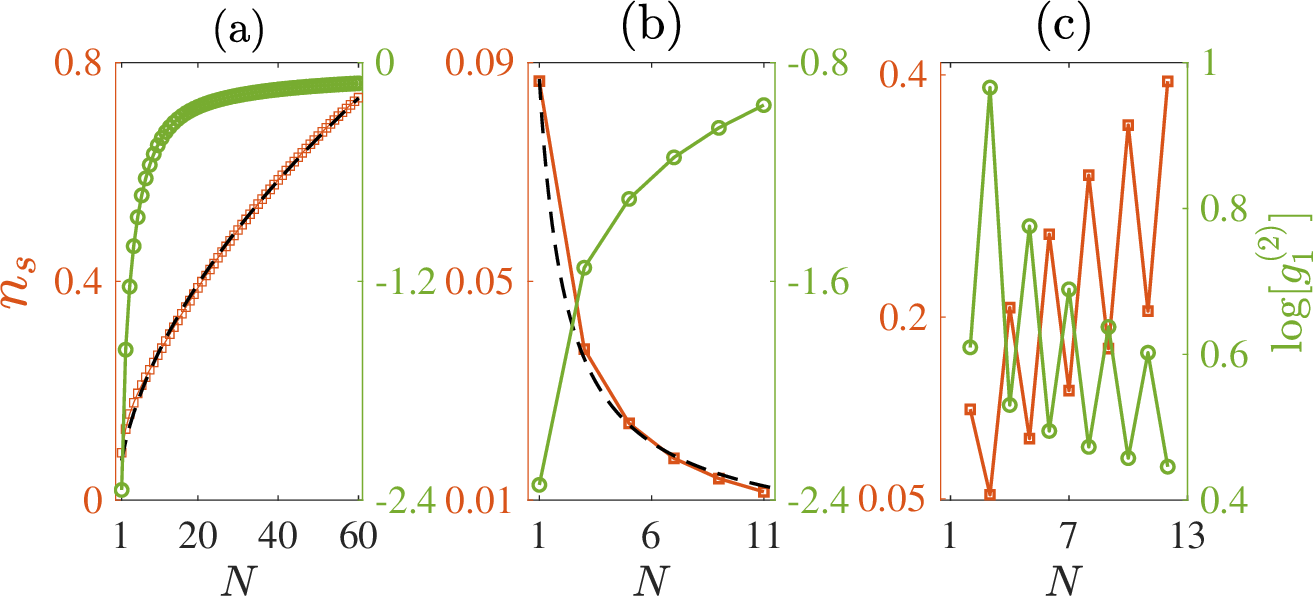} 
\caption{\label{fig3_scaling} $N$ dependence of $n_s$ ({red lines}) and $g^{(2)}_1$ ({green lines}) for (a)  $\phi=0$ and (b) $\phi=\pi$ at $\Delta=|\Delta_{1\pm}|$, and (c) $[\phi, \Delta]=[\pi, 0]$, respectively. The dashed lines indicate power-law fits: $n_s=0.07 N^{0.56}$ in (a) and $n_s=0.09 N^{-0.8}$ in (b).} 
\end{figure}

To characterize the subradiance-superradiance quantum phase transition, we introduce the order parameter ${\cal S}=n_s(N,\Delta,\phi)/[Nn_s(1,\Delta_s,\phi)]$ with $\Delta_s=g+\Delta-\Delta_{1\pm}$, such that ${\cal S}>1$ ($<1$) signals superradiance (subradiance). Clearly, $n_s(1,\Delta_s,\phi)$ also reaches its maximum at $\Delta=\Delta_{1\pm}$ for $\phi=\pi$ and  arbitrary $N$. Fig.~\ref{fig4_scaling}(a) shows the distribution of ${\cal S}$ on the $\Delta-\phi$ parameter plane. Due to photon blockade, the cavity always remains subradiant for $\Delta<\Delta_{1,+}$. Interestingly, the system undergoes a transition from subradiance to superradiance as $\Delta$ increases when $\phi/\pi<0.5$. In particular, we find ${\cal S}>3$ in the regime $\Delta/g \gg 1$ at small $\phi$, demonstrating enhanced superradiance. Conversely, even at large detuning, increasing $\phi$ drives a superradiance-to-subradiance transition due to destructive interference.

To further explore scaling, Fig.~\ref{fig4_scaling}(b) shows $n_s$ versus $N$ at $\phi=0$ for different $\Delta$. Notably, $n_s$ increases rapidly with $N$ and saturates to $N=(\Delta_{1\pm}/g)^2$ at single-photon resonance. For $N>(\Delta_{1\pm}/g)^2$, $n_s$ decreases monotonically, contrasting with the scaling $n_s \propto  N^2$ predicted by Eq.~(\ref{steady_state}). For $\Delta/g \gg 1$, we find that the steady-state photon number follows $n_s\sim N^{2.22}$ ($N^{2.07}$) for $\Delta/g=12$ ($20$). Thus, the quadratic $N^2$ scaling is recovered only in the strict far-dispersive limit, where mean-field description captures collective enhancement while neglecting quantum fluctuations~\cite{baumann2010dicke,PhysRevLett.112.143007,PhysRevResearch.5.013002}. 

Figures~\ref{fig3_scaling}(a) and \ref{fig3_scaling}(b) show $n_s$ and second-order correlation function $g^{2}_1(0)$ as a function of $N$ at $\Delta=|\Delta_{1\pm}|$ for different $\phi$. We find power-law scaling $n_s\sim N^{0.56}$ for $\phi=0$ and $n_s\sim N^{-0.8}$ for $\phi=\pi$. In both case, the single-photon purity decreases monotonically as $N$ increases. In the thermodynamic limit ($N\sim\infty$) and  $\phi=0$, cavity field evolves from single-photon state into coherent state with $g^{2}_1(0)$ to $1$, signaling a quantum-to-classical transition in which photon blockade is gradually destroyed. A striking feature appears at $\phi=\pi$, where photon emission exhibits a pronounced parity dependence on $N$. For even $N$, destructive interference suppresses $n_s$ entirely. This corresponds to antisymmetric bright states $|\Psi_{1,\pm}\rangle$ in Eq.~(\ref{eq_eigenvalue_single_ex}). For odd $N$, the imbalance $N_A-N_B=1$ breaks the symmetry, allowing residual photon emission. Remarkably, in both cases the scaling laws $n_s\sim N^{0.56}$ ($\phi=0$) and $n_s\sim N^{-0.8}$ ($\phi=\pi$) deviate strongly from mean-field prediction of Eq.~(\ref{steady_state}), demonstrating the crucial role of quantum fluctuations in collective cavity scattering. {Our results uncover a hierarchy of scaling laws from suppressed $1/N$ subradiance to enhanced $N^2$ superradiance controlled by phase-dependent interference.}

In Fig.~\ref{fig3_scaling}(c), we show $n_s$ and $g^{(2)}_1(0)$ at $\Delta=0$ and $\phi=\pi$. Although parity dependence persists, $n_s$ remains finite even for even $N$, in contrast to perfect destructive interference that appears at sidebands. Strikingly, the middle branch exhibits strong photon bunching ($g^{(2)}_1(0)\gg1$), suggesting a high-quality source of multiphoton states. We emphasize that photon emission with bunched statistics at middle branch yet studied in experiments~\cite{PhysRevLett.131.253603, PhysRevLett.130.173601}. Unlike superradiant coherent emission observed in the far-dispersive regime~\cite{PhysRevLett.131.253603}, our results reveal a transition from antibunched single-photon to bunched multiphoton emission under in atom-cavity resonance. Moreover, generating multiphoton emission on the middle branch requires pump atoms directly, rather than driven the cavity field~\cite{PhysRevLett.130.173601}. We also confirm that $n_s(\Delta=0)$ vanishes without external drive ($\Omega=0$). Finally, a small ratio $\gamma/\kappa$ is also essential for realizing nonclassical bunched emission in cavity-coupled alkaline-earth atom arrays.

\begin{figure}[ptb]
\includegraphics[width=0.49\textwidth]{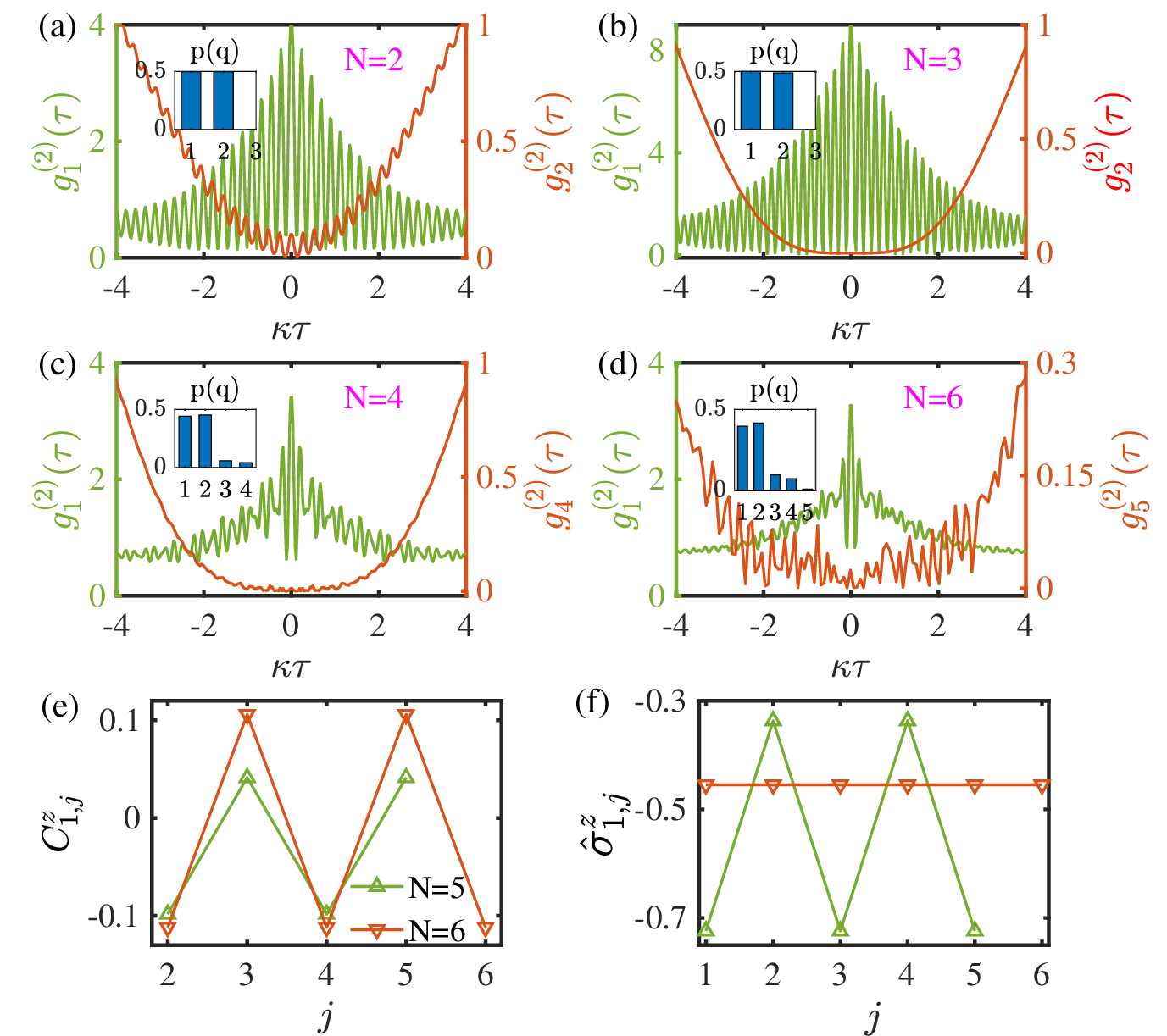} 
\caption{\label{fig_N_photon_bundle} (a)-(d) Time interval $\tau$ dependence of $g^{(2)}_1 (\tau)$ (solid lines) and  $g^{(2)}_n (\tau)$ (dashed lines)  for $N=2$, 3, 4, and 6, respectively. The inset displays the steady-state photon distribution $p(q)$. Spatial profiles of (e) correlations  $C^{z}_{1,j}$ and (f) spin magnetism  $\langle\hat{\sigma}^z_j\rangle$ as function of lattice site $j$ for different $N$. The other parameters are $\phi=\pi$ and $\Delta=0$.} 
\end{figure}

{\em N-photon bundle emission}.---To demonstrate nonclassical emission, we calculate interval dependence of correlation functions $g_1^{(2)}(\tau)$ and $g_n^{(2)}(\tau)$ for characterizing $n$-photon bundle generation. For $N=2$ and $3$ [Fig.~\ref{fig_N_photon_bundle}(a) and  Fig.~\ref{fig_N_photon_bundle}(b)], two-photon bundle emerge at middle branch ($\Delta=0$) with $\phi=\pi$, satisfying $g_1^{(2)}(0)>g_1^{(2)}(\tau)$ and $g_2^{(2)}(0)<g_2^{(2)}(\tau)$, signatures of photon bunching within a bundle and antibunching for separated bundles~\cite{PhysRevLett.117.203602,C2014Emitters, deng2021motional}. The associated correlation lifetimes significantly exceed $1/\kappa$, enabled by the long-lived alkaline-earth atoms with $\gamma/\kappa \ll 1$. For $N=4$ [Fig.~\ref{fig_N_photon_bundle}(c)], we observe a clearly resolved four-photon bundle, again exhibiting the characteristic correlation signatures $g_1^{(2)}(0)>g_1^{(2)}(\tau)$ and $g_4^{(2)}(0)<g_4^{(2)}(\tau)$. Remarkably, large $N$ further facilitates higher-order bundles [Fig.~\ref{fig_N_photon_bundle}(d)], with accessible order $n$ controlled by pump strength $\Omega$ and atom-cavity coupling $g$. Importantly, photon population remains sizable ($n_s>10^{-1}$) and increases substantially with $N$ [Fig.~ \ref{fig3_scaling}(c)]. 

To further quantify bundle emission, we evaluate steady-state photon-number distribution $p(q)={\rm tr}(q|q\rangle \langle q|\rho)/n_s$, which characterize the weight of $q$-photon states among total emitted photons. For $n$-photon bundle, $p(q)$ rapidly decreases and essentially vanishes for $q>n$, providing clear evidence of strong $(n+1)$-photon blockade. Unlike Mollow emission in far-dispersive regime~\cite{C2014Emitters}, our cavity-coupled reconfigurable atom arrays offer a practical route to high-quality multiphoton sources by completely suppressing single-photon excitation via interference, without requiring the strong-coupling limit and while mitigating detrimental heating or decoherence associated with a strong pump field.

 To elucidate microscopic origin of photon emission, we examine connected spin correlations $C^{z}_{i,j}=\langle \hat{\sigma}^z_{i}\hat{\sigma}^z_{j}\rangle - \langle \hat{\sigma}^z_{i}\rangle \langle\hat{\sigma}^z_{j}\rangle$ across all spin pairs $(i,j)$~\cite{emperauger2025tomonagaluttingerliquidbehaviorrydbergencoded,PhysRevLett.133.106901}. In Fig.~\ref{fig_N_photon_bundle}(e) and \ref{fig_N_photon_bundle}(f), we plot $C^{z}_{1,j}$ and spin magnetism $\langle\hat{\sigma}^z_j\rangle$ as a function of lattice site $j$.  A  pronounced alternation of positive (negative) correlations on even (odd) sites is observed, reflecting the intrinsic the configuration of AB sublattice. Notably, the correlation profiles $C^{z}_{1,j}$ is highly sensitive to the parity of $N$. In particular, even $N$ yields strong long-range order ($|C^{z}_{1,j}|\approx0.1$), favorable for multiphoton bundle formation. The alternating sign of $C^{z}_{1,j}$ reveals a $\pi$-phase shift between neighboring atoms arising from their half-integer wavelength separation. Moreover, the distance-independent amplitude of $C^{z}_{1,j}$ directly evidences the infinite-range nature of cavity-mediated interactions. Consistently, the spin magnetization $\langle\hat{\sigma}^z_j\rangle\approx-0.45$ remains homogeneous for even $N$, whereas odd $N$ exhibits a clear staggered magnetization.

In contrast, single-photon emission at $\Delta=\Delta_{1\pm}$ and $\phi=0$ shows correlation collapsing to $C^{z}_{1,j}\approx-0.01$ and nearly fully polarized spins $\langle\hat{\sigma}^z_j\rangle < -0.9$. This behavior is consistent with photon blockade: emission of the first photon suppresses subsequent excitations, producing high-purity single-photon states with vanishing inter-atomic correlations. Taken together, these results demonstrate that spin correlations in cavity-coupled arrays provide a powerful diagnostic for distinguishing emission regimes, ranging from pronounced single-photon blockade to the emergence of collective $n$-photon bundle.

{\em Conclusion}.---We have proposed a realistic and experimentally accessible scheme to realize tunable transitions between superradiant and subradiant emission in cavity-coupled atomic arrays. By systematically analyzing the power-law scaling of photon population with array size, we uncover that the hierarchy of photon correlations provides a direct signature of collective radiance transitions. The interplay of collective phase interference and atomic parity enables deterministic control over nonlinear quantum emission---from single-photon blockade to collective $n$-photon bundle generation. The observed parity- and phase-dependent spin correlations and magnetization provides a microscopic diagnostic for distinguishing distinct cooperative emission regimes. Our work establishes a new perspective on the many-body mechanisms underlying nonclassical light generation and presents a feasible route for engineering tunable quantum switch between high-quality single-photon to multiphoton sources in scalable cavity QEDs. Beyond atomic platforms, the proposed mechanism can be readily extended to cavity-coupled reconfigurable polar molecule arrays, where long-range dipolar interactions may drive novel superradiance and strongly correlated quantum emission~\cite{wang2025dark}. These results open promising directions for engineering hybrid quantum interfaces that exploit cooperative light-matter physics for quantum information processing and precision metrology~\cite{holland2023demand,bao2023dipolar}.

{\em Acknowledgments}.---This work was supported by the National Natural Science Foundation of China (Grant No.12374365, Grant No. 12274473, and Grant No. 12135018) and Guangdong University of Technology SPOE Seed Foundation (SF2024111504).

%\bibliographystyle{unsrt}
%\bibliography{references}

%apsrev4-2.bst 2019-01-14 (MD) hand-edited version of apsrev4-1.bst
%Control: key (0)
%Control: author (8) initials jnrlst
%Control: editor formatted (1) identically to author
%Control: production of article title (0) allowed
%Control: page (0) single
%Control: year (1) truncated
%Control: production of eprint (0) enabled
%

\end{document}